\documentclass[10pt]{article}
\usepackage{graphicx}
\usepackage{amsmath}

\input{tcilatex}

\begin{document}

\title{Poincar\'{e}'s relativistic Doppler-Fizeau formula}
\author{Dr. Yves Pierseaux \\
ULB ypiersea@ulb.ac.be}
\maketitle

\begin{abstract}
We deduce from Poincar\'{e}'s ellipsoidal wavefronts a relativistic
Doppler-Fizeau formula that is not the same as 1905 Einstein's one.
Longitudinally, Einstein's formula and Poincar\'{e}'s formula are the same.
Both formulas are compatible with relativistic time dilation. The question
of an experimental test is directly connected with the possibility or the
impossibility of directly measuring the transverse Doppler effect.
Hasselkamp's experiment (1978) becomes a crucial experiment because Poincar%
\'{e}'s relativistic kinematics predicts an expansion of space directly
connected with the Doppler-Fizeau effect for the remote objects.
\end{abstract}

\section{Lorentz transformation and Poincar\'{e}'s elliptical wavefront}

Let us consider a \textit{couple} of inertial systems K and K' in uniform
translation relative one another. A source of light is at rest O in K. What
is the image by LT in K', of a circular wavefront in K, emitted in $%
t^{\prime }=t=0$ by this source S? Poincar\'{e} writes LT with perfectly
spacetime symmetry (in x and t) \cite{8}:

\begin{equation}
x^{\prime }=k(x+\varepsilon t)\qquad \qquad \qquad y^{\prime }=y\qquad
\qquad t^{\prime }=k(t+\varepsilon x)
\end{equation}

Poincar\'{e}'s notations where $\varepsilon ,k$ correspond to
Einstein-Planck's notations $\beta ,\gamma $ given that, according to Poincar%
\'{e} in his 1905 work about the theory of relativity,\textit{\ ''I shall
choose the units of length and of time in such a way that the velocity of
light is equal to unity''\cite{4}.} For the easiness of present reader we
adopt standard notations $\beta ,\gamma $ but we keep Poincar\'{e}'s
spacetime units $c=1$'':

\begin{equation}
x^{\prime }=\gamma (x+\beta t)\qquad \qquad \qquad y^{\prime }=y\qquad
\qquad t^{\prime }=\gamma (t+\beta x)
\end{equation}

\bigskip Let us consider the relativistic invariant: 
\begin{equation}
x^{2}+y^{2}=r_{0}^{2}=t_{0}^{2}\qquad \qquad x^{\prime 2}+y^{^{\prime
}2}=r^{\prime 2}=t^{\prime 2}
\end{equation}

The object time $t=t_{0}$ is fixed in K (circular wavefront in K) but image
time \textit{t' } is not fixed (by LT) \textit{in} K'. We obtain with (2) $%
t^{\prime }=\gamma ^{-1}t_{0}+\beta x^{\prime }$, 
\begin{equation}
x^{\prime 2}+y^{^{\prime }2}=(\gamma ^{-1}t_{0}+\beta x^{\prime })^{2}\qquad 
\text{ou}\qquad (\gamma ^{-1}x^{\prime }-\beta t_{0})^{2}+y^{\prime
2}=t_{0}^{2}
\end{equation}
the Cartesian equation of an elongated ellipse\ in K'\`{e}the observer O' is
at the focus, \textbf{figure 1}). Poincar\'{e}'s ellipse is engraved in LT
and its physical meaning is the \textbf{relativity of simultaneity}: two
simultaneous events (\textit{different abscissa}) in K are not simultaneous
in K' \cite{1ter} \& \cite{7}: 
\begin{equation}
\Delta t=0\qquad \rightarrow \qquad \Delta t^{\prime }=\frac{t^{\prime
+}-t^{\prime -}}{2}\neq 0=\gamma \beta x
\end{equation}

$\Delta t^{\prime }$ is ''the gap of simultaneity'' between two opposite
events on the tront (\textbf{figure 1}). In order to have a circular image%
\footnote{%
If according to Einstein, the object (fixed time t) and the image (fixed
time t'), are both spherical within the two systems, then\ two simultaneous
events in K must be always also simultaneous in K'. \cite[paragraphe 3]{1}).
This is Einstein's convention of synchronisation in spherical waves ($\Delta
t=\Delta t^{\prime }=0)$\cite{7}$.$} of the wavefront we must fix the time
t' : 
\begin{equation}
(\Delta t^{\prime }=0)_{\text{Einstein}}  \tag{5bis}
\end{equation}
Poincar\'{e}'s ellipse in polar coordinates ($x^{\prime }=r^{\prime }\cos
\theta ^{\prime },$ $x^{\prime }=r^{\prime }\sin \theta ^{\prime }$with $%
\theta ^{\prime }$ as polar angle and F as pole) is :

\begin{equation}
r^{\prime }=\frac{r_{0}}{\gamma (1-\beta \cos \theta ^{\prime })}
\end{equation}

\bigskip With relativistic transformation of angle 
\begin{equation}
\cos \theta ^{\prime }=\frac{\cos \theta +\beta }{1+\beta \cos \theta }
\end{equation}

we rediscover polar LT for any point of the wavefront $(r,\theta $ or $%
t,\theta )$ :

\begin{equation}
r^{\prime }=\gamma r_{0}(1+\beta \cos \theta )\qquad \qquad t^{\prime
}=\gamma t_{0}(1+\beta \cos \theta )
\end{equation}

that are the radial or the temporal equation of an ellipse. The LT of a
circular wavefront into a elliptical wavefront has a 4-vectorial
interpretation: 
\begin{equation}
(r\cos \theta ,\text{ }r\sin \theta ,\text{ }0,\text{ }t)\qquad \rightarrow
\qquad (r^{\prime }\cos \theta ^{\prime },\text{ }r^{\prime }\sin \theta
^{\prime },\text{ }0,\text{ }t^{\prime })
\end{equation}

Light-like unprimed 4-vector (null interval) is transformed into\ light-like
primed 4-vector (null interval) (6) 
\begin{equation}
r^{\prime }\cos \theta ^{\prime }=\gamma (r\cos \theta +\beta t)\qquad
r^{\prime }\sin \theta ^{\prime }=r\sin \theta \qquad t^{\prime }=\gamma
(t+\beta r\cos \theta )
\end{equation}

with the invariance of the quadractic form (the interval of events, see
below):

\begin{equation}
\ r^{2}-t^{2}=0\qquad \rightarrow \qquad r^{\prime 2}-t^{\prime 2}=0
\end{equation}

The transformed (primed) 4-vector describes an \textit{isotropic ellipse} in
the meaning that the velocity of light (''one way'') is identical in all
directions within both systems because we have by construction \textbf{%
figure 1}):

\begin{equation}
\frac{r_{0}}{t_{0}}=\frac{r^{\prime +}}{t^{\prime +}}=\frac{r^{\prime -}}{%
t^{\prime -}}=c=1
\end{equation}

What is a wavefront in a relativistic sense (a ''spacetime'' wavefront)? The
invariant interval between the event ''emission'' $(t=t^{\prime }=0)$ and
any event on the wavefront $(t=t_{0},$ $r=r_{0}$ in K and $\ \ t^{\prime },$ 
$r^{\prime }$ in K'$)$ is null within both systems. So Minkowski's spacetime
4-vector (with null interval) is, in Poincar\'{e}'s kinematics, a \textbf{%
wavefront 4-vector (9)}\textit{. The elliptical image by LT is deduced from
a }\textbf{non-transversal }$\mathit{\ }t^{\prime }$\textit{\ \textbf{%
section }in Minkowski's cone\footnote{%
Poincar\'{e}'s spacetime metrics is not the same as Einstein-Minkowski's
spacetime metrics (\cite{4}) .}. }The essential fact that Poincar\'{e}'s
ellipse is engraved in LT can be seen with the inverse LT:

\begin{equation}
x=\gamma (x^{\prime }-\beta t^{\prime })\qquad \qquad \qquad y=y^{\prime
}\qquad \qquad t=\gamma (t^{\prime }-\beta x^{\prime })
\end{equation}

We obtain indeed immediately the temporal equation of the inverse ellipse (%
\textbf{figure 2}) (here polar equations, from 6 $r=t$ and $r^{\prime
}=t^{\prime }$):

\begin{equation}
\left[ t^{\prime }=\frac{t}{\gamma (1-\beta \cos \theta ^{\prime })}\right]
_{\text{direct ellipse}}\text{\qquad\ \ \ }\left[ t=\frac{t^{\prime }}{%
\gamma (1+\beta \cos \theta )}\right] _{\text{inverse ellipse}}\text{\ }
\end{equation}

Inverse elongated ellipse (14bis), with the source at the focus and the
observer O' at the centre, is Poincar\'{e}'s \textit{historical ellipse
(1906-1912)}. \ This ellipse explains, according to Poincar\'{e}, the null
result of Michelson experiment (the mean time back and forth is the same in
any direction, see Poincar\'{e}'s oscillator \cite{4}). Let us however here
focus the attention on the fact that the relation between t and t' is
exactly the same if we consider the ''source at rest'' (14) or the
''observer at rest''(14bis). Suppose now that the source emits a light
signal with a period of T seconds. Given that the relativistic kinematics is
a classical theory, the period of the emitted wave is also T seconds. The
periodical wave that consists of a series of elliptical wavefronts received
by the observer defines the standard configuration of a Doppler-Fizeau
effect (with the frequency $\nu =\frac{1}{T}$ and $\nu ^{\prime }=\frac{1}{%
T^{\prime }})$. In prerelativistic point of view the two cases are not
symmetrical. In relativistic (elliptical) point of view, the respective role
of the source and the observer are completely interchangeable (see Poincar%
\'{e}'s wavefront 4-vector). However, in order to have a rigorous deduction
of a Doppler-Fizeau formula (emission in a \textit{constant }direction), we
have to consider a plane wave (\cite{7}).

\section{Tangent to the ellipse, plane wave and Poincar\'{e}'s Doppler
formula}

Until now we considered one only source $S_{O}$ which emits at $t=t^{\prime
}=0$ a spherical wave with fixed time $t=t_{0}$ (figure 1)- Let us now
consider a second source $S_{\infty }$ at infinity at rest in K in the
direction $\theta $ (\textbf{figure 3}) which emits a plane wavefront.

Suppose that the considered plane wavefront (here a ''wave straight line''
at two space dimensions) be in O at the time $t=t^{\prime }=0$ (when $S_{O}$
emits a spherical wave). It will be \textbf{tangent} in $t=t_{0}$ to the
circular wavefront, emitted by $S_{1}$ (normalized $t_{0}=r_{0}=1$ \textbf{%
figure 3}). We note that the simultaneous events $P_{1}TP_{2}$ in K are no
longer simultaneous\textit{\ }($P_{1}^{\prime }T^{\prime }P_{2}^{\prime }$)
in K'. We deduce respectively (2) for the object front (''wave straight
line'' with angular coefficient $a=-\cot g$ $\theta $) and the image front
both following relations:

\begin{equation}
x\cos \theta +y\sin \theta =t_{0}\qquad \qquad y^{\prime }\sin \theta
^{\prime }+x^{\prime }\cos \theta ^{\prime }=t^{\prime }
\end{equation}

Exactly like the circular wavefronts (\'{e}quation 3), there are two
possibilities:

\textbf{If }$\mathbf{t}^{\prime }$\textbf{\ \ is not fixed (Poincar\'{e}, }%
spacetime image wavefront\textbf{),} we have by LT, $t^{\prime }=\gamma
(t+\beta x)=\gamma ^{-1}t+\beta x^{\prime }$, and therefore the primed
relation 15 is the equation of the tangent to the ellipse at the point $%
T^{\prime }$ (on which $P_{1}^{\prime }$ anf $P_{2}^{\prime }$ are situated$%
) $ :

\begin{equation}
y^{\prime }\sin \theta ^{\prime }+x^{\prime }(\cos \theta ^{\prime }-\beta
)=\gamma ^{-1}t
\end{equation}

The angular coefficient of Poincar\'{e}'s wavefront is $\theta ^{\prime }$:

\QTP{Body Math}
\begin{equation}
a_{poincar\acute{e}}^{\prime }=tg\alpha ^{\prime }=\frac{\beta -\cos \theta
^{\prime }}{\sin \theta ^{\prime }}
\end{equation}

\textit{\ If \ }$\mathbf{t}^{\prime }$\textbf{\ }\textit{\textbf{is fixed} 
\textbf{(Einstein, }}space image wavefront\footnote{%
The classical rigidity is incompatible with LT: a set of simultaneous events
cannot be remain by LT a set of simultaneous events. Einstein's rigid rods
are consistent with Einstein's rigid front wave.}\textit{\textbf{) }the
primed equation} 15 is the equation of the tangent to the circle (with
centre O' and radius $r_{T}^{\prime })$. The angular coeffecient of
Einstein's wavefront is: 
\begin{equation}
a_{einstein}^{\prime }=-\cot g\theta ^{\prime }
\end{equation}

Both formulas are the same for $\theta ^{\prime }=0.$ \textbf{Einstein's} 
\textbf{double transversality}\footnote{%
Einstein defined clearly the image of the wavefront in paragraph 7 of his
1905 paper \cite{1}: ''If we call the angle $\theta ^{\prime }$the angle
between \textbf{the wave-normal (direction of the ray)} and the ''direction
of movement''.''} (or \textbf{Einstein's double simultaneity}) involves for
the image front:\textit{\ } 
\begin{equation}
a^{\prime }=-\cot g\theta ^{\prime }\qquad \Leftrightarrow \qquad (\Delta
t^{\prime })_{front}=0
\end{equation}

This is consistent with Einstein's synchronisation (5bis). The phase $\Psi $
of a sinusoidal monochromatic plane wave ($\mathbf{A}$ is the amplitude) $%
\mathbf{A}=\mathbf{A}_{0}\sin \Psi $ ($\mathbf{A}^{\prime }=\mathbf{A}%
_{0}^{\prime }\sin \Psi ^{\prime })$ is defined by the \textbf{3}-\textbf{%
scalar product} $\mathbf{k.r}$ ($\mathbf{k}^{\prime }.\mathbf{r}^{\prime }%
\mathbf{)}$ with the frequency $\nu =\frac{\omega }{2\pi }$, the wave vector 
$\mathbf{k}=\frac{2\pi }{\lambda ^{\prime }}\mathbf{1}_{n}$ with $\mathbf{1}%
_{n}$ the unit vector normal to the front ( $\mathbf{k}^{\prime }=\frac{2\pi 
}{\lambda ^{\prime }}\mathbf{1}_{n^{\prime }},\nu ^{\prime }=\frac{\omega
^{\prime }}{2\pi }):$ 
\begin{equation}
\omega t-\mathbf{k.r=}\text{ }\Psi =\omega ^{\prime }t^{\prime }-\mathbf{k}%
^{\prime }\mathbf{.r}^{\prime }=\Psi ^{\prime }\qquad \Leftrightarrow \qquad 
\mathbf{A}.\mathbf{k=A}^{\prime }.\mathbf{k}^{\prime }=0=Ak\cos \phi
=A^{\prime }k^{\prime }\cos \phi ^{\prime }
\end{equation}

This is a \textbf{Galilean invariant} (t is fixed on the object front and t'
is fixed on the image front) in the meaning where the angle (\textbf{figure 3%
}) $\phi =\phi ^{\prime }=90%
{{}^\circ}%
$ is not altered by Galilean transformation (GT).

\begin{equation}
\mathbf{k.r}\ \qquad \underrightarrow{TG}\ \qquad \mathbf{k}^{\prime }.%
\mathbf{r}^{\prime }\mathbf{\ \ \ \ \ \ \ \ }\text{ou\qquad }\mathbf{A}\perp 
\mathbf{k}\ \qquad \underrightarrow{TG}\ \qquad \mathbf{A}^{\prime }\perp 
\mathbf{k}^{\prime })  \tag{20bis}
\end{equation}
But LT changes (17) this angle $\phi =90%
{{}^\circ}%
$ into $\phi ^{\prime }$ in the following way (\textbf{figure 3}): 
\begin{equation}
\tan \phi ^{\prime }=\tan (\alpha ^{\prime }-\theta ^{\prime })=\frac{\beta
\cos \theta ^{\prime }-1}{\beta \sin \theta ^{\prime }}  \tag{17bis}
\end{equation}

The right angle $\phi =\phi ^{\prime }=90%
{{}^\circ}%
$ is conserved \textit{if and only if }the propagation is purely
longitudinal $\sin \theta ^{\prime }=0$ or ''everything happens as if $\beta
=0"$, in other words exactly as in Einstein's synchronisation in spherical
waves (5bis, note1). Einstein considers \cite[paragraphe 7]{1} that the
Galilean invariant (20) is \textit{by definition} a Lorentz invariant.
Einstein's definition of the invariance of the phase is therefore based on a 
\textbf{4-scalar product} between the 4-vector position $(t,$ $\mathbf{r})$
and another 4-vector with null norm $(\omega ,$ $\mathbf{k}):$

\begin{equation}
(\omega ,\text{ }\mathbf{k}).(t,\text{ }\mathbf{r})=\omega t-\mathbf{%
k.r=\Psi \qquad }\text{\ }\nu (t-x\cos \theta -y\sin \theta )=\nu ^{\prime
}(t^{\prime }-x^{\prime }\cos \theta ^{\prime }-y^{\prime }\sin \theta
^{\prime })
\end{equation}

On the basis of Einstein's wave 4-vector (with two space dimensions)
Einstein deduces from the covariance of scalar product $\mathbf{k.r}$ a
Doppler-Fizeau formula for a monochromatic sinusoidal wave plane: 
\begin{equation}
(\frac{\nu ^{\prime }}{\nu })_{\text{Einstein}}=\gamma (1-\beta \cos \theta )
\end{equation}

Einstein's extension of prerelativistic concept of wave \textbf{3-vector}
(double transversality) into a wave \textbf{4-vector} $(\omega ,$ $\mathbf{k}%
)$ is a very serious problem ($t^{\prime }$ is fixed on the front but $%
r^{\prime }$ is not fixed) because the LT is not compatible with the double
transversality (double simultaneity).

Suppose now that the source $S_{2}$ (at the infinity, \textbf{figure 3})
emits a series of plane wavefronts (here a train of ''wave straight line'')
at the same\textit{\ rythm} as the spherical wavefronts ($\nu =\frac{1}{T}=%
\frac{1}{t})$. Let us construct a Lorentz invariant phase $\Phi $ on the
only basis of Lorentz invariance of the interval, $(t-r)(t+r)=(t^{\prime
}-r_{T}^{\prime })(t^{\prime }+r^{\prime }),$ between the event $t=t^{\prime
}=0$ and any event on the plane front (\cite{7}). With the tangency point T'
of the front to the ellipse (image of T, \textbf{figure 3}) we have $%
(t-r_{T})(t+r_{T})=(t^{\prime }-r_{T^{\prime }}^{\prime })(t^{\prime
}+r_{T^{\prime }}^{\prime })$:

\begin{equation}
\nu (t-r_{T})\text{ }\mathbf{=}\text{ }\Phi =\nu ^{\prime }(t^{\prime
}-r_{T^{\prime }}^{\prime })=\Phi ^{\prime }\qquad \text{or \ \ \ \ \ \ }%
\omega t-kr_{T}=\Phi =\ \omega ^{\prime }t^{\prime }-k^{\prime }r_{T^{\prime
}}^{\prime }=\Phi ^{\prime }
\end{equation}

Poincar\'{e}'s phase invariant $\Phi =\Phi ^{\prime }$, which is constructed
with fundamental quadratic invariant (3), is necessarily a solution of the
second order wave equation of electromagnetism (\cite{5} \& \cite{6}). The
invariance of the phase for the wavefront is no longer coupled with the
(prerelativistic) fixing of the time on the wavefront in each system. We
immediately deduce from (23) a relativistic Doppler-Fizeau formula: 
\begin{equation}
(\frac{\nu ^{\prime }}{\nu })_{Poincar\acute{e}}=\gamma (1-\beta \cos \theta
^{\prime })
\end{equation}

Let us call this formula, deduced from Poincar\'{e}'s ellipse,
''Poincar\'{e}'s Doppler-Fizeau formula''. \textbf{This is not the same as
Einstein's one} (22) because the proper angle $\theta $ is replaced with
improper angle $\theta ^{\prime }$ (see conclusion 29 for \textbf{%
experimental tests}).For the longitudinal propagation, the formulas are the
same:

\begin{equation}
(\frac{\nu ^{\prime }}{\nu })_{\text{Einstein-Poincar\'{e}}}=\sqrt{\frac{%
1-\beta }{1+\beta }=}\gamma (1-\beta )
\end{equation}

Poincar\'{e}'s formula can be immediately deduced from (14), with $\nu =%
\frac{1}{t}$ et $\nu ^{\prime }=\frac{1}{t^{\prime }},$ and therefore for
polar LT (8), supposes the existence of a \textbf{wave 4-vector} \textit{%
directly proportional} to the wavefront 4-vector defined in the paragraph 1 (%
$c\neq 1)$ (9)

\begin{equation}
(\lambda \cos \theta ,\text{ }\lambda \sin \theta ,\text{ }0\text{, }%
cT)_{poincar\acute{e}}\qquad \qquad (\frac{1}{\lambda }\cos \theta ,\frac{1}{%
\lambda }\sin \theta ,\text{ }0,\text{ }\frac{\nu }{c})_{einstein}
\end{equation}
Poincar\'{e}'s 4-vector involves that the frequency (24) is transformed by
LT \textit{like the inverse of a the time} $\nu =\frac{1}{t}$ et $\nu
^{\prime }=\frac{1}{t^{\prime }}$ whilst Einstein's four vector\footnote{%
In Einstein-Minkowski's kinematics, the light has a double 4-vectorial repr%
\'{e}sentation : the Minkowski's light like four-vector and Einstein's wave
4-vector. This latter is inseparable du photon d'Einstein, d\'{e}fini en
jauge transverse \cite{5}.} involves that frequency (22) is transformed by
LT \textit{like the time }(see de Broglie's works). Poincar\'{e}'s
proportionality (9-26) involves that this Doppler-Fizeau formula for remote
objects ($S_{\infty })$ is not independant of the structure of space-time
(Poincar\'{e}'s \textbf{expansion of space} for remote objects \cite{7}). We
showed \cite{6} that Poincar\'{e}'s theory of wavefront is compatible with
electromagnetic theory of waves (in particular with the transversality of
the primed electric field to the primed direction of propagation).

\section{Conclusion: Doppler formulas and experimental tests}

\bigskip In order to experimentally test both Doppler formulas, which are
both compatible with the relativistic dilation of time \cite{4}, we need
length waves (or frequencies) emitted by moving atoms with respect to the
observer (the relativistic transformation of the energy is not in question
here). Mandelberg describes the SODS (''Second Order Doppler-Shift''):

\begin{quote}
A moving beam of radiating hydrogen atoms with velocities ranging up 2.8*10$%
^{8}cm/s$ had been viewed from the incoming and the outgoing directions
simultaneously. Averaging wavelength measurements of a particular spectral
line for the two observations give a measurement of the quadratic shift...
The wavelength of the light emitted from an oncoming atom of velocity v
viewed at a small angle $\theta ^{\prime }$\ ( near zero) to the beam
direction by an observer stationary in the laboratory reference frame is
given by (1) where $\theta ^{\prime }$\ is measured in the laboratory frame
and $\lambda $\ is the wavelenght observed in a reference frame at rest with
respect to the radiating atom. The $\beta \cos \theta $\ term in the
numerator is the first-order Doppler shift. the subscript B indicates a 
\textit{Doppler shift to the blue}. The denominator contains the quadratic
shift; the appproximation expression is valid in this experiment because of
the low velocity and the experimental uncertainty. Similarly the wavelength
of the light emitted by the backward direction, i.e. as seen by a laboratory
fixed observer looking at a receding atom is given by (2) \cite{2ter}

\begin{equation}
\lambda _{B}^{\prime }=\lambda \frac{1-\frac{\text{v}}{c}\cos \theta
^{\prime }}{\sqrt{1-\beta ^{2}}}=\lambda (1-\beta \cos \theta ^{\prime }+%
\frac{1}{2}\beta ^{2})\text{ \ }(1)\qquad \lambda _{R}^{\prime }=\lambda 
\frac{1+\frac{\text{v}}{c}\cos \theta ^{\prime }}{\sqrt{1-\beta ^{2}}}%
=\lambda (1+\beta \cos \theta ^{\prime }+\frac{1}{2}\beta ^{2})\text{ \ }(2)
\end{equation}

the subscript R indicated a \textit{red Doppler shift}. The average of these
two wavelenghts 
\begin{equation}
\lambda _{Q}=\frac{\lambda _{B}^{\prime }+\lambda _{R}^{\prime }}{2}=\lambda
(1+\frac{1}{2}\beta ^{2})
\end{equation}
\ equals to the wavelength which would be observed at right angles to the
beam corresponding $\theta =\frac{\pi }{2}$\ in previous equations. There
are many reasons why a perpendicular observation of the beam is not
feasible...
\end{quote}

This kind of experiment measures in fact longitudinally (for $\theta
^{\prime }\sim \theta \sim 0)$ the factor 1/2 with respect to the second
order term $\beta ^{2}$ (Mandelberg measures $0,498\pm 0.025)$ $.$
Mandelberg defines a arithmetic average with the angle $\theta ^{\prime }$.
With an arithmetic average based on the angle $\theta ,$ we obtain exactly
the same result with Poincar\'{e}'s formula $(\frac{\lambda ^{\prime }}{%
\lambda })_{Poincar\acute{e}}=\gamma (1-\frac{\text{v}}{c}\cos \theta
)\rightarrow \lambda _{Q}=\frac{\lambda _{B}^{\prime }+\lambda _{R}^{\prime }%
}{2}=\lambda (1+\frac{1}{2}\beta ^{2})$. Longitudinally Einstein's formula
and Poincar\'{e}'s formula are the same (25). 

Given that SODS experiments directly  measure (with the primed angle of the
observer in the laboratory $\theta ^{\prime })$ the length waves, we can
write respectively ''blueshift'' and ''redshift'' \ Einstein's and Poincar%
\'{e}'s formulas with $\theta ^{\prime }\leq 90%
{{}^\circ}%
$ (22-24) in the following\ way: 
\begin{equation}
(\frac{\lambda _{B}^{\prime }}{\lambda })_{\text{Einstein}}=\gamma (1-\frac{%
\text{v}}{c}\cos \theta ^{\prime })\qquad (\frac{\lambda _{B}^{\prime }}{%
\lambda })_{\text{Poincar\'{e}}}=\frac{1}{\gamma (1+\frac{\text{v}}{c}\cos
\theta ^{\prime })}
\end{equation}

\begin{equation}
(\frac{\lambda _{R}^{\prime }}{\lambda })_{\text{Einstein}}=\gamma (1+\frac{%
\text{v}}{c}\cos \theta ^{\prime })\qquad (\frac{\lambda _{R}^{\prime }}{%
\lambda })_{\text{Poincar\'{e}}}=\frac{1}{\gamma (1-\frac{\text{v}}{c}\cos
\theta ^{\prime })}
\end{equation}

Let us note that, without the relativistic factor $\gamma $, we find again
both forms of classical formulas respectively of blueshift and reshift (the
forms are inverted if we use unprimed proper angles $\theta $). According to
Mandelberg \cite{2}, the measurement of transversal effect '' ($\theta
^{\prime }=\frac{\pi }{2})$ is not feasible\footnote{%
Let us remark that the experimental point of view defined the transversality
with observer angle whilst the theoretical point of view defined the
transversality with the proper angle of the source. The question of the
definition of transversality becomes very important with Poincar\'{e}'s
formula.}''. Most of physicists have quoted the direct observation of the
transversal effect to be extremely difficult or even impossible. This is the
reason why most of experiments are on the Ives-Stilwell longitudinal
configuration. At the first sight it seems impossible \cite{1bis} to
experimentally distinguish between Poincar\'{e}'s and Einstein's formula.
However, in 1979, Hasselkamp \cite{2bis} introduces an experiment that
directly measures the quadratic relativistic effect in the vicinity of the
angle $\theta ^{\prime }=90%
{{}^\circ}%
$ . He rejects the argument according to which the Doppler-broadening of the
line (which is a consequence of the finite opening angle of the optical
system) could be much larger than the second order Doppler-shift). Hasselkamp%
\textbf{\ }specifies ''the mechanical could only be performed to an accuracy
in the real observation angle $90%
{{}^\circ}%
<\theta ^{\prime }<91%
{{}^\circ}%
$. And also ''The real angle of observation is $90,5%
{{}^\circ}%
$''. This is important for the discussion because Hasselkamp measures a
redshift and if we use the blueshift formula (29 with $90,5%
{{}^\circ}%
$) the quadratic effect is compatible with both formulas $\cos \theta
^{\prime }=-\frac{\text{v}}{c}:$ $(\frac{\lambda ^{\prime }}{\lambda }%
)_{Poincar\acute{e}}=\frac{1}{\gamma (1-\frac{\text{v}^{2}}{c^{2}})}=\gamma $
if the magnitude of the velocities $\frac{\text{v}}{c}\simeq 0.01$ but not
compatible with Poincar\'{e}'s formula for $\frac{\text{v}}{c}\simeq 0.03.$
\ The function of velocity seems to confirm Einstein's formula. But the
question is not completely clarified because according the figure 1 \cite
{2bis} , the domain of measure $90%
{{}^\circ}%
<\theta ^{\prime }<91%
{{}^\circ}%
$ is clearly a domain of \textit{incoming} direction of atoms. So we must
use the blueshift formulas (29, $\theta ^{\prime }\leq 90%
{{}^\circ}%
$ ) between the angle 90$%
{{}^\circ}%
$ and $89,5%
{{}^\circ}%
$ \ for which Einstein's formula gives respectively a redshift and a
blueshift\cite{3}. In order to experimentally make the difference between
Einstein's formula and Poincar\'{e}'s formula we have to measure within a
symmetrical domain around the right angle $89,5%
{{}^\circ}%
<\theta ^{\prime }<91%
{{}^\circ}%
.$

\end{document}